# Electronic structure of BaFeO$_3$: an abinitio DFT study

## Hong-Jian Feng, Fa-Min Liu

Department of Physics, School of Sciences, Beijing University of Aeronautics & Astronautics, Beijing 100083, P. R. China

**Abstract:** First principles calculations were performed to study the ground state electronic properties of BaFeO$_3$ (BFO) within the density functional theory (DFT). Adopting generalized gradient approximation (GGA) exchange and correlation functional and Vosko-Wilk-Nusair correlation energy functional interpolation, we have systematically conducted the band structure, density of states and electronic distribution along different crystalline planes. Calculating results show that band gap in the majority spin band structure and band gap in the minority spin band structure were found to be 2.7012 eV and 0.6867 eV respectively. Up-spin Fe t$_{2g}$ were fully occupied and down-spin Fe e$_g$ were empty. Moreover, the up-spin Fe e$_g$ and down-spin Fe t$_{2g}$ were partially occupied near the Fermi energy, leading to a finite density of states. The Fe$^{4+}$-O-Fe$^{4+}$ plane superexchange coupling should rearrange the magnetic order to make the ferromagnetic characteristic being possible, moreover the tetragonal displacement along the c axis could induce the perovskites materials to acquire ferroelectric property. These reasons could lead to the fact that the tetragonal phase BFO could be a potential multiferroics while it was produced under the very experimental conditions. The charge density along different crystalline planes were illustrated to show that strong covalent bonding between O and Fe can be used to investigate the exchange coupling, and this strong hybridization may further increase the superexchange coupling to enhance the magnetic ordering.

**Introduction**

Multiferroics materials have attracted more interesting in recent years in terms of possessing magnetic order and ferroelectricity in the same phase. Since 1960s various review articles have systematically classified their properties and behaviors [1-5]. Previous research show that the magnetoelectric property may originate from the coupling between ferroelectric and ferromagnetic order parameters [6-10]. Even now the driven mechanism of the magnetoelectric multiferroics has not been discovered totally. More investigations on the theoretical methods need to be performed to push forward the progress on this sort of fascinating materials. BaFeO$_3$ (BFO) has been expected to be a potential magnetoelectrics due to the ferromagnetic ordering of Fe$^{4+}$ ions, and it was reported to be hexagonal structure in the bulk form. Eiichi Taketani and his collaborators have predicted that the BFO thin films with a tetragonal crystalline structure might be a potential candidate for multiferroics [11]. Therefore we might speculate the BFO with tetragonal crystalline structure in the bulk form should be an ideal candidate for multiferroics as it was produced with this special tetragonal perovskites structure under a certain experimental condition.

First principles calculations have been

recently used to describe the electronic structure of materials, as far as perovskites type materials are concerned. One of the first theoretical investigations in BTO ferroelectric transition has been done by Cohen and Krakouer in the early 1990s using the first principles calculations based on density functional theory (DFT) with local density approximation (LDA) method [12-13]. In this paper, electronic properties of BFO in bulk form with tetragonal crystalline structure have been systematically investigated by using first principles calculation based on density functional theory (DFT) within local density approximation (LDA) method. And we mainly focused on the calculations of the band structure, density of states (DOS) and the charge density along different crystalline planes within the DFT exchange and generalized gradient approximation (GGA) correlation functional to provide comprehensive understanding on the forthcoming experiments.

## 2. Computational details:

The calculation was separated into three steps. Firstly, the lattice constant optimization was performed with respect to the lowest energy. The initial lattice constant (a = 0.39904 nm, c = 0.40689 nm) was provided, which has been derived from the average value calculated from the X-ray diffraction pattern shown in Fig.1.

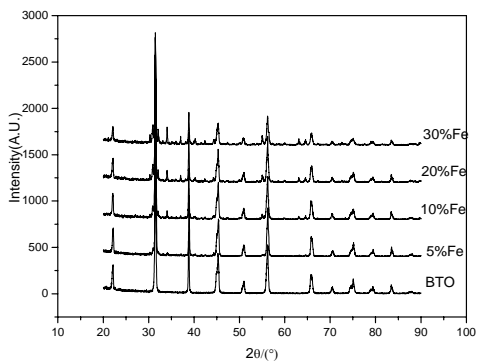

Fig.1 X-ray diffraction patterns of different Fe doping BaTiO$_3$

Secondly, we conducted the self-consistent calculation, and the cell was constructed in Fig.2.

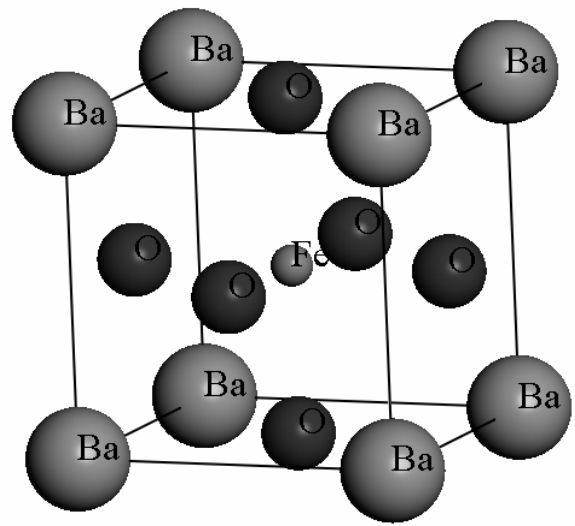

Fig. 2 The diagram of tetragonal structural BaFeO$_3$

The corresponding space group was p4 mm. During relaxation the cell shape and volume was fixed. In the third step, we performed the non-self-consistent calculation. In our DFT computations, we used Vosko - Wilk - Nusair correlation energy functional and generalized gradient approximation (GGA) exchange and correlation functional as suggested by Perdew and Wang (PWGGA). The reciprocal space integration was performed by sampling of the Brillouin Zone with the 11×11×11 Monkorst-Pack net. The high symmetry points in the reciprocal space were selected to describe the band structure of BFO. Plane-wave functions were used as basis sets, and a plane-wave cutoff energy of 500 eV was employed throughout. It has shown that the results are well converged at this cutoff. Spin-orbit interaction was excluded in our calculations due to its weak influences in 3d elements. Ba 5s, Ba 5p, Fe 3s, Fe 3p, Fe 3d, O 2s and O 2p electrons have been treated as valence state.

**3 Results and discussion**

From the band structure for majority spin in Fig. 3, one can see that the quasi-flat band positioned at -17 eV originated from the O 2s states, and those positioned at -11 eV originated from the Ba 5p states. These results can be approved from the partial DOS of Ba and O atoms respectively. The band in the vicinity of the Fermi level was formed by the hybridization of O 2p and Fe 3d states. Due to the p-d overlap in this portion, the charge redistribution was completed to form the compound BFO. The topmost in the majority spin band structure was derived from the Ba 6s states, and this point can also be proved by the corresponding orbital resolved DOS of Ba. The band gap calculated was 2.7012 eV. Moreover the topmost point in the valence band (VB) is localized in the high symmetry point M along the Brillouin zone, and the lowest point in the conduction band (CB) is found in the high symmetry point Γ along the Brillouin zone. The band structure of the minority spin was depicted in Fig.4, in which no significant difference was observed. However, the band gap here is decreased to 0.6867 eV, and a relative small value compared with the majority spin, which was caused by the shift of down-spin Fe $t_{2g}$. The corresponding highest point in VB and lowest point in CB have not been changed accordingly.

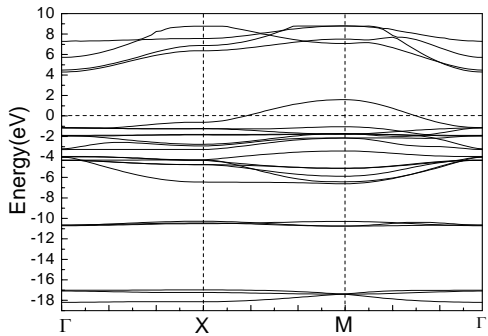

Fig. 3 The majority spin band structure calculated for $BaFeO_3$

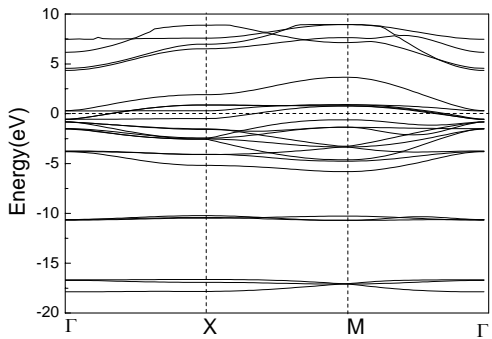

Fig. 4 The minority spin band structure calculated for $BaFeO_3$

From the spin-resolved total DOS it can be seen that the spin up and spin down states have not equaled in the upper and lower panel shown in the Fig. 5, which was the total DOS of BFO.

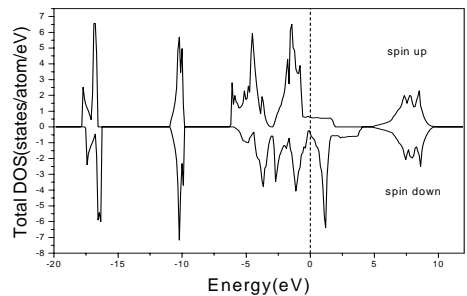

Fig. 5 The total DOS Vs energy

The asymmetry in the vicinity of the Fermi level attributed to the spin ordering of the Fe atom which might become the origin of the ferromagnetic property in this special structure material. Up-spin Fe $t_{2g}$ were fully occupied and down-spin Fe $t_{2g}$ were partially filled, leading to a finite density of states at the Fermi energy.

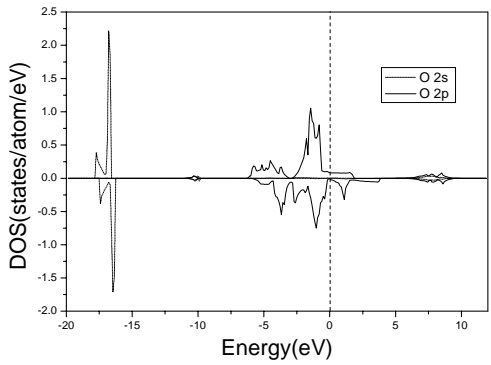

Fig. 6 Spin-resolved DOS of O Vs energy

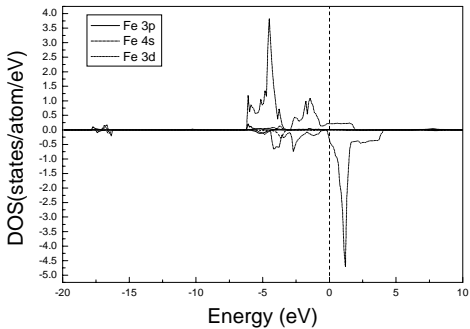

Fig. 7 Spin-resolved DOS of Fe Vs energy

However, up-spin Fe $e_g$ were partly filled, and down-spin Fe $e_g$ were fully empty. The spin-resolved DOS of O and Fe atom were illustrated in Fig.6 and Fig. 7, and the O 2s states at -17 eV was confirmed the lowest lying band in the band structure again. The asymmetry was also observed in the vicinity of the Fermi energy, and it is assumed that this might be caused by the interaction between the octahedral oxygen and the Fe atom. The up-spin Fe $e_g$ and down-spin Fe $t_{2g}$ were partially occupied near the Fermi energy in the Fig. 7 leading to a finite DOS.

Generally speaking, the magnetization caused by the $Fe^{4+}(d^4-d^4)$ type exchange coupling depend on the species and the arrangement of the magnetic ions, and the $Fe^{4+}$-O-$Fe^{4+}$ plane superexchange coupling should rearrange the magnetic order to make the ferromagnetic property become possible. Moreover, the tetragonal displacement along the c axis could induce the perovskites materials to acquire ferroelectric property. Hence these two reasons could provide the fact that the tetragonal phase BFO might be a potential multiferroics while it was produced under the very experimental conditions.

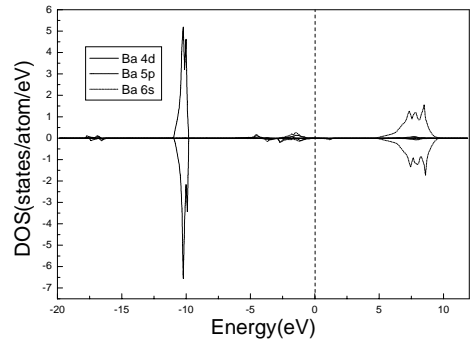

Fig. 8 Spin-resolved DOS of Ba Vs energy

In Fig. 8 the spin-resolved DOS of Ba assured that the band positioned at -11 eV was derived from the Ba 5p states and the highest region in the band structure could also found its counterpart directly.

The charge density along different crystalline planes has been shown in Fig. 9, Fig.10, Fig.11 and Fig. 12 respectively.

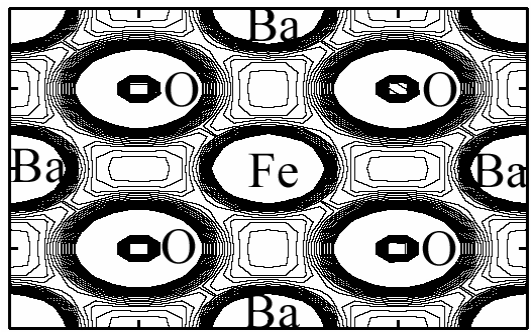

Fig. 9 The charge density distribution along

(001) plane.

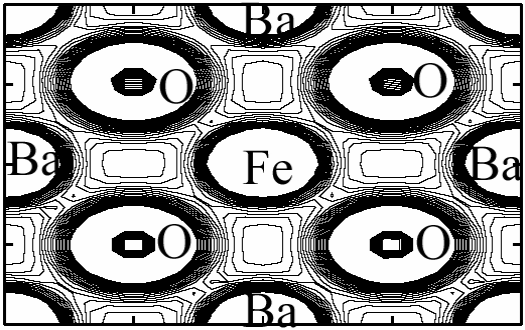

Fig. 10 The charge density distribution along (100) plane.

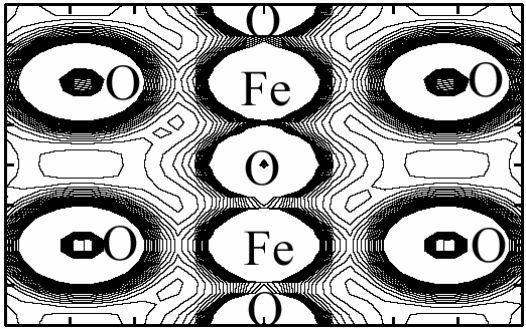

Fig. 11 The charge density distribution along (110) plane.

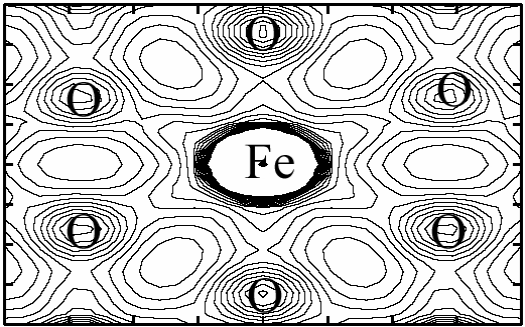

Fig. 12 The charge density distribution along (111) plane.

The charge density distribution along the (001) crystalline plane and (100) crystalline plane are shown in Fig.9 and Fig.10, respectively. There was no sharp difference between these two charge density maps due to the p4mm group symmetry. The strong covalent bonding between O and Fe atom can be investigated from the maps, and this results further proved that the p-d overlap between O 2p and Fe 3d orbitals. The charge redistribution along (110) crystalline plane and (111) crystalline plane has been conducted in Fig.11 and Fig.12. The strong covalent bonding between Fe and O atoms have been affirmed again by the charge map along the (110) crystalline plane. Even in the (111) crystalline plane the hybridization could be observed due to the strong p-d overlap. Therefore this strong hybridization may further increase the superexchange coupling to enhance the magnetic ordering. Although most of papers published previous have made great endeavour to discover the charge distribution of perovskites compounds, they had not projected the charge density onto different crystalline planes. Our investigation aims show a new approach to study the electronic distribution with respect to the space group symmetry.

## 4. Conclusions

We have systemically performed the band structure, density of states and electronic distribution along different crystalline planes by using the first principles calculation based on the density functional theory within the local density approximation scheme. The calculating results show that the band gap in the majority spin band structure was found to be 2.7012 eV, whereas the one for the minority spin was decreased to be only 0.6867 eV, causing by the shift of down-spin Fe $t_{2g}$ in the positive direction of the x coordinate. From the spin-resolved total DOS it can be seen that the spin up and spin down states have not equaled in the upper and lower panel, which agree well with the conclusion in the band structure. The quasi-flat band positioned at -17 eV and -11 eV was attributed to the O 2s states and Ba 5p states, respectively. The up-spin Fe $e_g$ and down-spin Fe $t_{2g}$ were partially occupied near the Fermi energy leading to a finite density of states, and the $Fe^{4+}$-O-$Fe^{4+}$ plane superexchange coupling should rearrange the magnetic order to make the ferromagnetic property become possible. So it can be inferred that the tetragonal phase BFO might be a potential multiferroics while it was displaced along certain crystalline direction. The charge density along different crystalline planes shows that strong covalent bonding between O and Fe can be used to investigate the exchange coupling.

**Acknowledgements:**

This work was supported by the Aeronautical Science foundation of China (2003ZG51069).